\documentclass[preprint2,11pt]{aastex}

\newcommand\puncspace{\ifmmode\,\else{\ifcat.\C{\if.\C\else\if,\C\else\if?\C\else%
\if:\C\else\if;\C\else\if-\C\else\if)\C\else\if/\C\else\if]\C\else\if'\C%
\else\space\fi\fi\fi\fi\fi\fi\fi\fi\fi\fi}%
\else\if\empty\C\else\if\space\C\else\space\fi\fi\fi}\fi}
\newcommand\SP{\let\\=\empty\futurelet\C\puncspace}

\newcommand{\Msun}{{\rm M}_\sun\SP}

\newcommand{\saxpsr}{SAX J1808.4-3658\SP}

\slugcomment{Submittted to ApJ Letters: 2 February 2000}

\shorttitle{Relativistic Effects in SAX~J1808.4-3658}
\shortauthors{Ford}

\begin{document}

\title{Relativistic Effects in the Pulse Profile of the 2.5
msec X-Ray Pulsar SAX~J1808.4-3658}

\author{Eric C. Ford}

\affil{Astronomical Institute, ``Anton Pannekoek'', University of Amsterdam}
\affil{Kruislaan 403, 1098 SJ Amsterdam, The Netherlands}
\email{ecford@astro.uva.nl}


\begin{abstract}

We analyze the properties of the pulsed emission from the accreting
millisecond pulsar \saxpsr in observations of its April 1998 outburst
by the Rossi X-Ray Timing Explorer. Pulse phase spectroscopy shows
that the emission evolves from a hard spectrum (power law with photon
index $2.39\pm0.06$) to a soft spectrum (index $3.39\pm0.24 $). This
softening is also observable as a phase lag in the fundamental of
low-energy photons with respect to high-energy photons. We show that
this lag is roughly constant over ten days of the outburst.  We fit
these data with a model where the pulse emission is from a hot spot on
the rotating neutron star and the flux as a function of phase is
calculated including the effects of general relativity.  The
energy-dependent lags are very well described by this model. The
harder spectra at earlier phases (as the spot approaches) are the
result of larger Doppler boosting factors which are important for this
fast pulsar.

\end{abstract}


\section{Introduction}

Strong X-ray pulsations with a 2.49 msec period were discovered from
\saxpsr in an April 1998 observation with the Rossi X-ray Timing
Explorer \citep{wk98}. The pulsar is a member of an accreting binary
system with an orbital period of 2.01 hour and a low-mass companion
\citep{cm98}. Though similar to other low-mass X-ray binaries in its
timing and spectral properties \citep[e.g.][]{wk99,hs98}, \saxpsr is
unique for its X-ray pulsations. No other such binary has shown
coherent pulsations in its persistent flux despite of careful searches
\citep[][and references therein]{vaughan94a}.

As such, \saxpsr is the fastest rotating accreting neutron star.
If the pulsations are due to modulated emission from one hot spot on
the neutron star surface, the 2.49 msec period corresponds to an
equatorial speed of approximately 0.1c. With these high speeds,
\saxpsr offers an excellent system for studying relativistic effects.

One such effect may be the observed lag of low-energy photons relative
to high-energy photons in the pulse discovered by
\citet{cmt98}. \citet{cmt98} suggest that the lags are due to
Comptonization in a relatively cool surrounding medium. Alternatively,
the lags may be the result of a relativistic effect: the high-energy
photons are preferentially emitted at earlier phases due to Doppler
boosting along the line of sight. This possibility was suggested for
the similar lags in the 549~Hz oscillations in an X-ray burst of
Aql~X-1 \citep{ford99}, where a simple model showed that the delays
roughly match those expected.

In the following, we present new measurements of the pulsed emission
from \saxpsr. We show that the energy-dependent phase lags are
equivalent to a hardening pulse profile. We model this behavior in
terms of a hot spot on the neutron star, including relativistic
effects.

\section{Observations \& Analysis}

We have used publicly available data from the proportional counter
array (PCA) on board RXTE in an `event' mode with high time resolution
(122$\mu$sec) and high energy resolution (64 channels). The
observations occurred from April 10 1998 to May 7 1998, when the source
was in outburst.

We generate folded lightcurves in each PCA channel. This is
accomplished with the fasebin tool in FTOOLS version 4.2, which
applies all known XTE clock corrections and corrects photon arrival
times to the solar system barycenter using the JPL DE-200 ephemeris,
yielding a timing accuracy of several $\mu$sec (much less than the
phase binning used here). As a check, we have applied this method to
Crab pulsar data and the results are identical to \citet{pravdo97}.
To produce pulse profiles in the neutron star rest frame, we use the
\saxpsr orbital ephemeris found by \citet{cm98}. An example folded
lightcurve is shown in Figure~\ref{lc} (top) for the observation of
April 18 1998 14:05:40 to April 19 1998 00:51:44 UTC.

To study the energy spectra at each phase bin, we take the rates at
pulse minimum and subtract it from the rest of the data at other
phases. This effectively accomplishes background subtraction and
eliminates the unpulsed emission which we do not wish to
consider. Note, however, that the pulsed emission may have some
contribution even at the pulse minimum and this subtraction scheme
represents only a best approximation to the true pulse emission. We
generate detector response matrices appropriate to the observation
date and data mode using pcarsp v2.38, and use XSPEC v.10.0 to fit
model spectra.

We fit the spectrum here with a simple powerlaw function. Though the
function itself is not meant to be a physical description, the
powerlaw index provides a good measure of the spectral hardness.  Fits
in each phase bin have reduced $\chi^2$ of 0.7 to 1.6. Including an
interstellar absorption does not substantially affect these results.
The powerlaw index clearly increases through the pulse phase
(Figure~\ref{lc}, bottom), i.e. spectrum evolves from hard to soft.

We also fit the profiles of the folded lightcurves in each channel
using Fourier functions at the fundamental frequency and its
harmonics. From these fits we determine the phase lag in each channel
relative to the fits in some baseline channel range. Results for the
18 April 1998 observation are shown in Figure~\ref{lags} as solid
symbols. Note that negative numbers indicate that high-energy photons
precede low-energy photons. We are also able to measure lags in the
first harmonic, and find that they are opposite in sign to the
fundamental, i.e. low-energy photons precede high-energy photons in
the first harmonic. No lags are measurable in the other harmonics.

Another way to measure energy-dependent phase lags is by Fourier
cross-correlation analysis. This is the method used for \saxpsr by
\citet{cmt98} and for other timing signals as well, e.g. kilohertz
QPOs \citep[e.g.][]{kaaret99a}.  For a description of cross
correlation analysis see e.g. \citet{vaughan94b}.  From the PCA event
mode data we calculate Fourier spectra in various channel ranges with
Nyquist frequency of 2048 Hz and resolution of 0.25 Hz. We then
calculate cross spectra defined as $C(j) = X_1^{*}(j)X_2(j)$, where
$X$ are the complex Fourier coefficients for two energy bands at the
pulsar frequency $\nu_j$. The phase lag between the two energy bands
is given by the argument of $C$. We measure all phase delays relative
to the unbinned channels range 5 to 8, i.e. 1.83 to 3.27 keV for 5
detector units in PCA gain epoch 3 (April 15 1996 to March 22
1999). The results for the 18 April 1998 observation, are shown in
Figure~\ref{lags} by the open symbols.  The phase lags are consistent
with those calculated from the lightcurve fitting. From the
cross-correlation spectra we are not able to measure lags in the much
weaker harmonics.


We have calculated phase lag spectra also for other RXTE observations
during the April 1998 outburst. These spectra are similar to that in
Figure~\ref{lags}. To quantify the trends, we compute an average phase
delay, $\phi_{avg}$, over all energies for each observation.  We also
fit a broken powerlaw function to each phase delay spectrum:
$\phi=E^{-\alpha}$ below a break energy, $E_{b}$, and
$\phi=\phi_{max}$ above this energy.  Figure~\ref{lags_v_time} shows
the quantities $\phi_{avg}$, $E_{b}$ and $\phi_{max}$ versus the
time of each observation.

There is a clear connection between the results of the two analyses
presented here. The phase resolved spectroscopy shows that the
spectrum softens and correspondingly the peak of the pulse profile
appears slightly earlier in phase for higher energies
(Figure~\ref{lc}). The method of measuring phase delays shows the same
behavior: higher-energy photons preferentially lead lower-energy
photons in the fundamental and the magnitude of this phase delay
increases with energy (Figure~\ref{lags}). In the following we discuss
a model that can account for the phase delays/spectral softening
measured here.

\section{Model}

We calculate the expected luminosity as a function of phase in a
manner similar to \citet{pfc83,stroh92} but including Doppler effects
\citep{cs89} and time of flight delays \citep{fkp86}.  This treatment
is based on a Schwarzschild metric, where the photon trajectories are
completely determined by the compactness, $R/M$. The predicted
luminosity as a function of phase from our code matches the results of
\citet{pfc83} and \citet{cs89} for the various choices of parameters.
\citet{brr00} recently developed a model for pulse profiles using a
slightly different approach.

The parameters in the model are the speed at the equator of the
neutron star, $v$, the mass of the neutron star, $M$, the compactness,
$R/M$, the angular size of the cap, $\alpha$, and the viewing angles,
$\beta$ (the angle between the rotation axis and the cap center) and
$\gamma$ (rotation axis to line of sight). Another ingredient is the
emission from the spot, which we take as isotropic and isothermal. The
spectrum of energy emitted from the spot is another important
input. Note that if the emitted spectrum is power law like, the
observed spectrum will not evolve with phase, since Doppler
transformation preserves the shape \citep[see][]{cs89}. The intrinsic
spectrum must therefore have some shape that transforms to match the
observed hardening and phase lags; we use a blackbody spectrum with
temperature $kT_0$.

A fit from the model is shown in Figure~\ref{lags}. This fit uses the
following model parameters: $R/M=5$, $M=1.8\Msun$, $kT_0=0.6$ keV,
$v=0.1$, $\beta=\gamma=10^{\rm o}$, $\alpha=10^{\rm o}$.  To derive
count rates, we use table models in XSPEC and the appropriate response
files as discussed above.  The fit for this single set of parameters
is good; we find $\chi^2=5.3$ from the Fourier cross-correlation data
or a reduced $\chi^2$ of 0.8 for all the parameters fixed.

A full exploration of the parameter space of the model is beyond the
scope of this letter. We note the following trends, however.  The
magnitude of the phase lags depends sensitively on $v$ and
$kT_0$. Larger delays result from higher speeds, because the pulses
become increasingly asymmetric due to Doppler boosting. This asymmetry
is also energy dependent, so there is a dependence on $kT_0$ as well
\citep[see][]{cs89}.  The lags also depend on $\beta$ and $\gamma$,
especially at higher energies where there is the turn-over noted
above. The phase lags are less sensitive to $R/M$ and $\alpha$.  We
have tested the assumption of isotropic emission and found that the
phase lags depend only weakly on the beaming.

\section{Discussion}

\begin{figure}[h]
\begin{center}
\epsscale{0.7}
\plotone{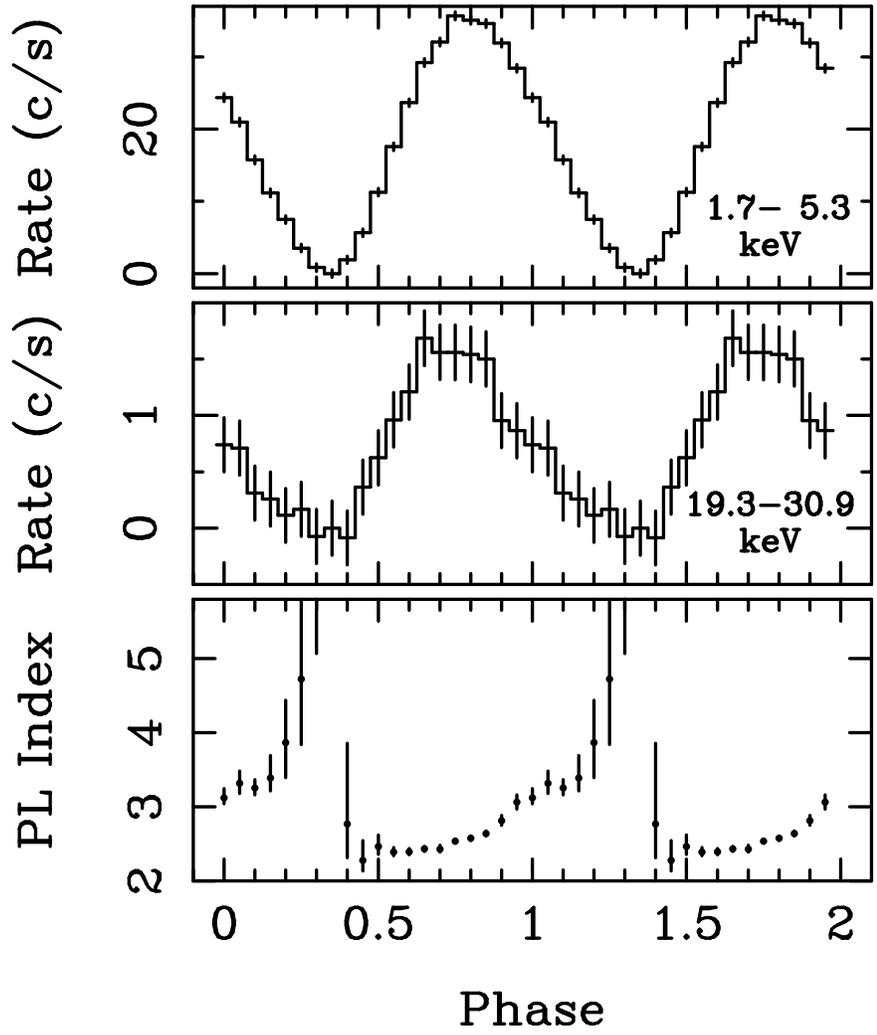}  
\caption{Folded lightcurve in example low (top) and high (middle)
energy bands and index of the power law of spectral fit (bottom). The
profile is repeated twice for clarity.}
\label{lc}
\end{center}
\end{figure}

\begin{figure}[h]
\begin{center}
\epsscale{0.7}
\plotone{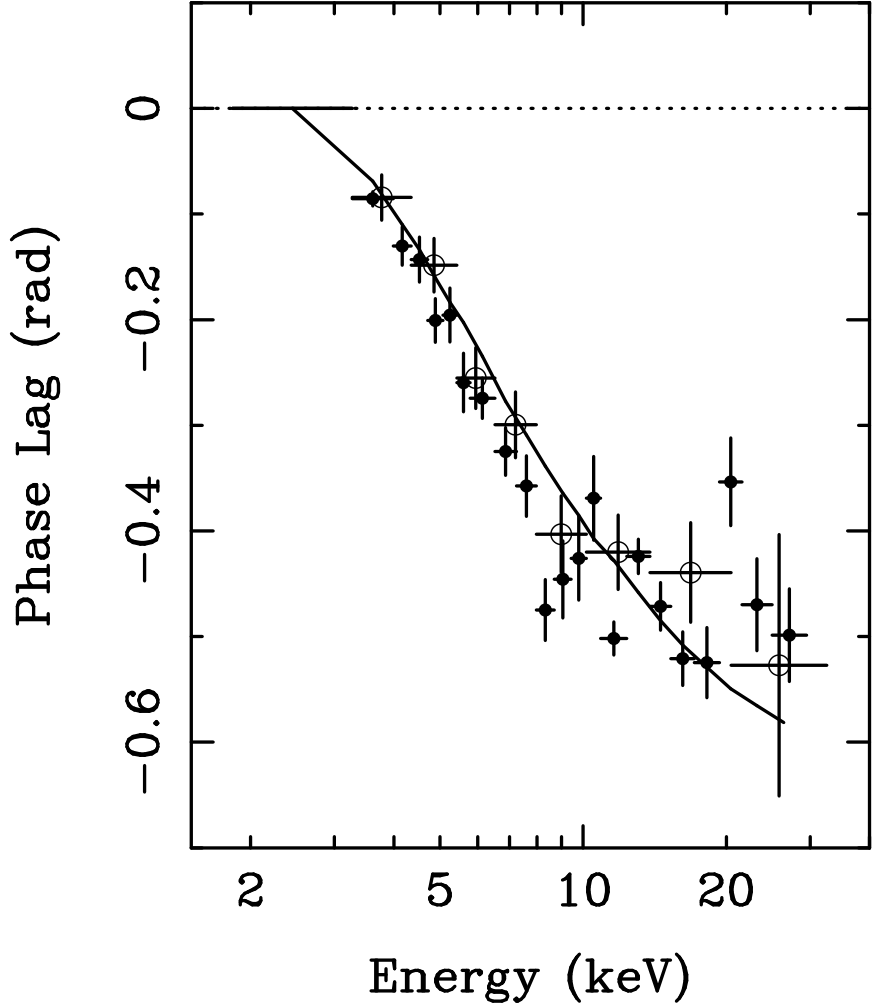}  

\caption{Phase delays in the fundamental of SAX~J1808.4-3658 relative
to the 1.83 - 3.27 keV band.  A negative number indicates that
high-energy photons precede low-energy photons. Solid symbols indicate
measurements from fitting the folded lightcurves, open symbols are
data from Fourier cross-correlation analysis. The line shows the model
for parameters $R/M=5$, $M=1.8{\rm M}_\sun$, $kT_0=0.6$ keV, $v=0.1$
c, and $\beta=\gamma=10^{\rm o}$ as described in the text.}

\label{lags}
\end{center}
\end{figure}

The pulsed emission from \saxpsr evolves through its phase from a
relatively hard to a soft spectrum, as shown by our phase resolved
spectroscopy (Figure~\ref{lc}). This evolution can also be thought of,
and measured as, an energy-dependent phase lag in the fundamental
(Figure~\ref{lags}), i.e. higher-energy photons emerging earlier in
phase than lower-energy photons.

\begin{figure}[h]
\begin{center}
\epsscale{0.8}
\plotone{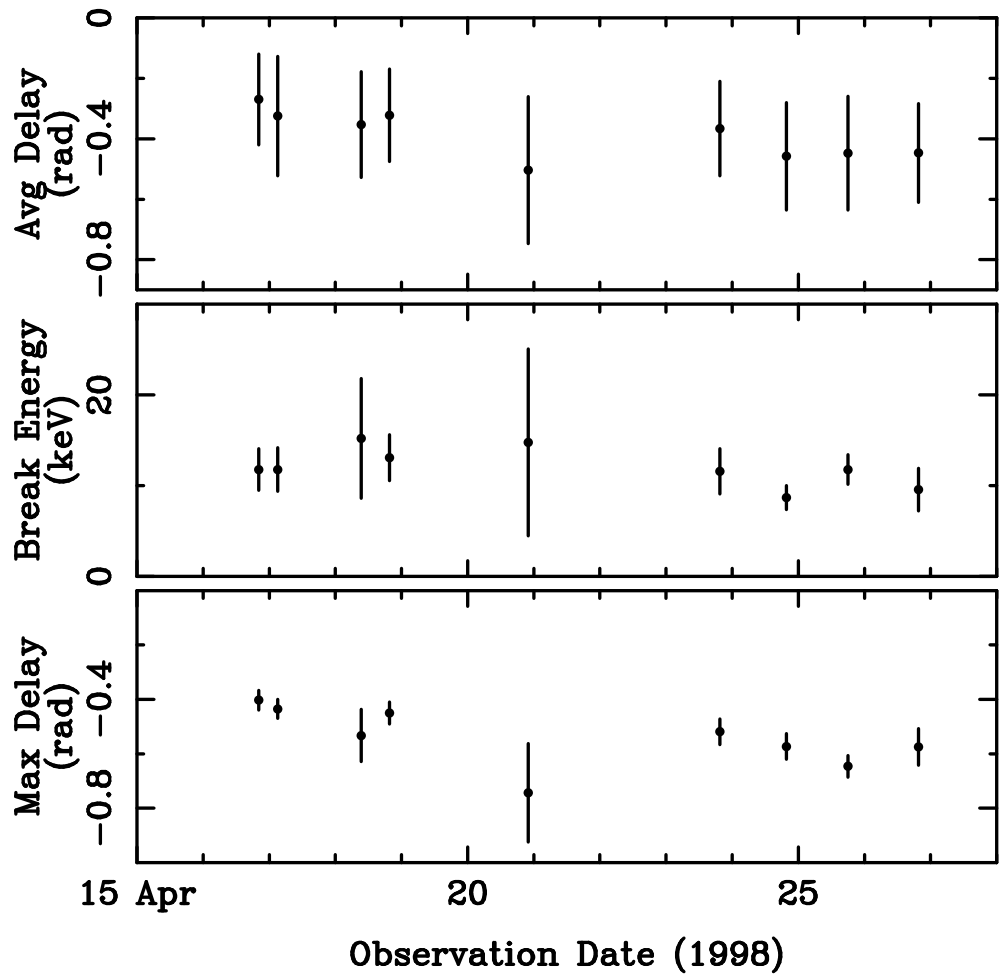}  
\caption{Parameters of the Fourier phase delay spectra for each
observation. Plotted are average phase delay over all energies (top),
the break energy (middle) and the maximum phase delay (bottom).}
\label{lags_v_time}
\end{center}
\end{figure}

We have applied a model to the data which consists of a hot spot on
the rotating neutron star under a general relativistic treatment. The
dominant effect accounting for energy-dependent delays is Doppler
boosting, the larger boosting factors at earlier phases giving harder
spectra. The model fits the data quite well.  The model also provides
a stable mechanism for generating the phase delays, which meshes with
the fact that the characteristic delays remain stable in time to
within 25\% (Figure~\ref{lags_v_time}) even as there is a factor of
two decrease in the X-ray flux, a possible tracer of accretion
rate. As noted in \citet{ford99}, in addition to explaining the
energy-dependent phase lags in \saxpsr, this mechanism may also
account for the lags in the burst oscillations of Aql X-1
\citep{ford99} and kilohertz QPOs \citep{vaughan98,kaaret99a}. Phase
resolved spectroscopy has not yet been possible in these signals.

The model offers a new means of measuring the neutron star mass and
radius, a notoriously difficult problem in accreting binary
systems. The radius is directly related to $v$, the equatorial speed
($v=\Omega_{\rm spin} R$). The fits also depend on the model
parameters $M$ and $R/M$, though the lag spectra are less sensitive to
these quantities.  The model could be independently constrained by
future optical spectroscopy of the companion which could provide
information on $M$ and the viewing angles, $\alpha$ and $\beta$.

\acknowledgments

We acknowledge stimulating discussions with the participants of the
August 1999 Aspen workshop on Relativity where an early version of
this work was presented. We thank Michiel van der Klis, Mariano
M\'{e}ndez and Luigi Stella for helpful discussions. This work was
supported by NWO Spinoza grant 08-0 to E.P.J.van den Heuvel, by the
Netherlands Organization for Scientific Research (NWO) under contract
number 614-51-002, and by the Netherlands Researchschool for Astronomy
(NOVA). This research has made use of data obtained through the High
Energy Astrophysics Science Archive Research Center Online Service,
provided by the NASA/Goddard Space Flight Center.


\onecolumn

\end{document}